# Non-thermal nature of photo-induced insulator-to-metal transition in $NbO_2$

Rakesh Rana[1*], J. Michael Klopf[2], Jörg Grenzer[1], Harald Schneider[1], Manfred Helm[1,3], and Alexej Pashkin[1**]

[1]*Institute of Ion Beam Physics and Material Research, Helmholtz-Zentrum-Dresden-Rossendorf, 01328 Dresden, Germany*

[2]*Institute of Radiation Physics, Helmholtz-Zentrum Dresden-Rossendorf, 01328 Dresden, Germany*

[3]*Institute of Applied Physics, Technische Universität Dresden, 01062 Dresden, Germany*

**Abstract:**

We study the photo-induced metallization process in niobium dioxide $NbO_2$. This compound undergoes the thermal insulator-to-metal transition at the remarkably high temperature of 1080 K. Our optical pump – terahertz probe measurements reveal the ultrafast switching of the film on a sub-picosecond timescale and the formation of a metastable metallic phase when the incident pump fluence exceeds the threshold of ~10 mJ/cm$^2$. Remarkably, this threshold value corresponds to the deposited energy which is capable of heating $NbO_2$ only up to 790 K, thus, evidencing the non-thermal character of the photo-induced insulator-to-metal transition. We also observe an enhanced formation of the metallic phase above the second threshold of ~17.5 mJ/cm$^2$ which corresponds to the onset of the thermal switching. The transient optical conductivity in the metastable phase can be modeled using the Drude-Smith model confirming its metallic character. The present observation of non-thermal transition in $NbO_2$ can serve as an important test bed for understanding photo-induced phenomena in strongly correlated oxides.

[*]r.rana@hzdr.de

[**]a.pashkin@hzdr.de

Understanding the subtleties of the insulator-to-metal transition (IMT) in strongly correlated oxides is an important challenge in condensed matter physics. One of the prime examples is the IMT at 340 K in $VO_2$ which is a $3d^1$ transition-metal oxide. The orbital-assisted collaborative effort of Mott physics and a Peierls transition is suggested as the IMT mechanism [1, 2]. Ultrafast time-resolved studies have demonstrated the photo-induced switching of $VO_2$ from the insulating phase into a metastable metallic state on a sub-100 fs time scale when the incident pump fluence exceeds a certain threshold value [3, 4]. While the ultrafast photoemission data [5] show evidence for a purely electronic character of the metallization immediately after the photoexcitation, it is known that the long-living metallic phase can be induced only when the absorbed pump energy exceeds the threshold necessary to heat the sample above $T_c$ [6]. Thus, the enthalpy change in the lattice subsystem across the photo-induced phase transition may play an important role in the formation of the metastable metallic phase. On the other hand, the results of other terahertz - [7] and electron diffraction [8] studies of $VO_2$ indicate that in a certain range of pump fluences, the metallization can be induced without the structural transition into the high-temperature rutile phase. A better understanding of this complex ultrafast switching behavior calls for a search for related oxide systems where the photo-induced metallization can be clearly identified as a non-thermal phase transition.

$NbO_2$ is a $4d^1$ system isovalent with $VO_2$ which, according to literature, transforms from a regular rutile structure into a tetragonal distorted rutile structure below the transition temperature $T_c$ = 1080 K [9]. This fact is related to a stronger bond existing between local Nb dimers due to a better overlap of 4d orbitals causing larger orbital splitting between $d_{||}$ (occupied) states and the $e^{\pi}{}_g$ (unoccupied) states, thus ensuring a robust Peierls effect in the insulating phase [10,11]. The formation of Nb dimers in the insulating phase has a quasi-one-dimensional character along the c-axis and is attributed to the instability due softening of the phonon mode at *P*-wave vector, $q_p = \left(\frac{1}{4}, \frac{1}{4}, \frac{1}{2}\right)$ [12]. However, the exact nature of the phase transition is still debated. Using density functional theory (DFT) it was speculated that the dimerization alone rather than electron correlations is responsible for the opening of the band gap [13]. Recent DFT and cluster-dynamical mean-field theory calculations stressed the role of electronic correlations which are also relevant for $NbO_2$, (although to a lesser extent than in $VO_2$) and thus should be taken into account along with the Peierls distortion, for a complete description in the insulating phase [14]. $NbO_2$ serves as an ideal choice for understanding the IMT in $VO_2$ as there are a few interesting similarities: both belong to the $d^1$ family of compounds and the (V-V or Nb-Nb) dimers in the insulating phase exhibit tilting with respect to the rutile c-axis [15]. Furthermore, in the high-temperature rutile phase, $VO_2$ and $NbO_2$ crystals have nearly the same axial ratio of c/a = 0.625. This value becomes slightly larger for monoclinic $VO_2$ and smaller for the tetragonal $NbO_2$, however, the two structures seem to be nearly degenerate near the phase transition [15]. Recently the ultrafast switching of $NbO_2$ films on femtosecond time scale has been demonstrated and compared with the response of $VO_2$ films using near-infrared pump-probe measurements [16].

Here we present a time-resolved optical pump– terahertz (THz) probe (OPTP) spectroscopic study of a $NbO_2$ thin film and demonstrate that the metastable metallic phase can be induced for pump fluences which heat the system to temperatures far below $T_c$. Our results identify a large range of pump fluences between 10 and 17.5 mJ/cm$^2$ over which a purely non-thermal metallization can be achieved.

The 190 nm thick epitaxial $NbO_2$ film was grown on (111) $MgAl_2O_4$ substrate using pulsed laser deposition technique (supplementary information s1) [17,18]. Time-resolved measurements were performed using a femtosecond Ti:sapphire regenerative amplifier delivering 2 mJ pulses with 35 fs FWHM duration at a 1 kHz repetition rate at the central wavelength of ~ 800 nm (1.55 eV). The near-infrared beam was split such that 90% served as a pump, while the remaining 10% of the beam was utilized to generate and electro-optically sample the THz probe pulses using a 2 mm ZnTe emitter and 0.4 mm GaP detector crystals. An aperture of 600 μm was placed over the sample in order to ensure probing in a homogeneously photoexcited region. The pump spot diameter was approximately 1.5 mm. The measured penetration depth for our $NbO_2$ film at the pump wavelength is 60 nm.

The sample was excited using the optical pump with the photon energy of 1.55 eV which is well above the direct (1.3 eV) and indirect (0.7 eV) band gaps of $NbO_2$ [13,18]. This process leads to the creation of a large number of non-equilibrium free carriers and a strong increase of optical conductivity at THz frequencies. As a result, the THz transmission through the thin film demonstrates a drop which is roughly proportional to the concentration of the free carriers. The dynamics of this process can be monitored from the temporal evolution of the differential THz transmission $\Delta T/T$ depicted in figure 1(a). Around zero delay time the pump-probe signal experiences an initial steep drop to negative values on the timescale limited only by the temporal resolution of our OPTP setup. For low pump fluences, it is followed by the fast relaxation within 1 ps and the recovery of the initial insulating state. This behavior is similar to the behavior of $VO_2$ described in previous studies and it is usually assigned to the recombination (localization) of the photoexcited charge carriers [4,7,19]. We find that as the fluence is increased above 10 mJ/cm$^2$, ΔT/T does not vanish after the initial relaxation and remains finite even after tens of picoseconds (see figure 1(a) and the inset). This remanence of ΔT/T suggests the formation of metastable metallic domains [19]. Further, we also noted that the pump fluence has a significant impact at earlier timescales (below 1ps). It may be seen that normalized ΔT/T exhibits a more rapid decrease (say around 0.5 ps) as depicted in the inset of figure 1 (b), which suggests that the fast relaxation speeds up with the increasing pump fluence and exhibits a saturation behavior after a certain threshold fluence.

To gain further insight into the relaxation mechanism, the ΔT/T was fitted with a biexponential fitting function, $A_f \exp(-t/\tau_f) + A_s \exp(-t/\tau_s) + A_{offset}$, where $\tau_f$ ($A_f$) and $\tau_s$ ($A_s$) represent the time constant (and amplitude) of the fast and the slow processes, respectively and $A_{offset}$ describes the constant offset. The latter value $A_{offset}$ represents a long-lived photo-induced metallic response. In figure 2 (a), the value of $A_{offset}$ exhibits a distinctive dependence on pump fluence with a typical dual threshold behavior. To capture this, extrapolation of the linear fits to the values of $A_{offset}$ in this plot gives two characteristic threshold fluences: $\Phi_{T1}$ ~ 10 mJ/cm$^2$ and $\Phi_{T2}$ ~ 17.5 mJ/cm$^2$. Above $\Phi_{T1}$, the non-vanishing $A_{offset}$ starts to increase moderately, and above $\Phi_{T2}$ the slope becomes steeper with a rapid increase. It is noteworthy that the value of the first threshold fluence $\Phi_{T1}$ for $NbO_2$ is nearly twice the corresponding value observed for $VO_2$ [4,6]. For $NbO_2$ the contribution from the slow dynamics ($A_s$) is an order in magnitude less than the faster dynamics ($A_f$) as seen in figure 2 (c). The ratio of $A_f/A_s$ increases from 4 to 10 with increasing fluence up to 24.7 mJ/cm$^2$ thereby suggesting that a large number of photoexcited carriers becomes localized within about 1 ps after photoexcitation. The slower dynamics captured by $\tau_s$ is interpreted as the thermalization of high energy molecular modes heated by the recombination of the quasiparticles, shows a decrease from 6 ps and saturates to a value of 3 ps above $\Phi_{T1}$. The faster dynamics characterized by $\tau_f$ also exhibits a steady decrease from 600 fs to 400 fs and remains saturated at this value with no effect of pump fluence above $\Phi_{T1}$. This saturation behavior of $\tau_f$ and $\tau_s$ above 10 mJ/cm$^2$ cannot be merely explained by carrier photoexcitation across the bandgap and may be an indication of the onset of the ultrafast IMT in the $NbO_2$ film.

As discussed earlier the effective thickness of our $NbO_2$ film for the pump excitation was found to be ~ 60 nm for the wavelength of ~ 800nm. Estimating the effective pump pulse energy deposited, a pump fluence of 1 mJ/cm$^2$ corresponds to an enthalpy change of ~ 3.52 kJ/mol in the case of $NbO_2$ [20]. Knowing this we can compare the temperature (***T****) of the photoexcited $NbO_2$ layer immediately after pumping with previously reported equilibrium enthalpy changes [20]. The enthalpy difference between the room temperature insulating phase of $NbO_2$ and its metallic phase at 1080 K is about 62.4 kJ/mol including the transition latent heat of 3.4 kJ/mol [20]. The threshold fluences $\Phi_{T1}$, $\Phi_{T2}$ and the upper limit of measured fluence (24.7mJ/cm$^2$) correspond to ***T**** - values of 790 K, 1080 K, and 1390 K in the photoexcited layer, respectively (figure 2(a)). The first threshold $\Phi_{T1}$ nearly corresponds to the energy of 33 kJ/mol for the $NbO_2$. The second threshold $\Phi_{T2}$ corresponds to the $T_c$ where ***T**** starts to exceed Tc and a part of the thin film may be switched also thermally. Naturally, the activation of the additional switching mechanism leads to the more rapid growth of the metallic phase fraction within the insulating phase and is well captured in figure 2(a) by a rapidly increasing $A_{offset}$ above a fluence of $\Phi_{T2}$. In order to check the purely nonthermal IMT the excitation regime between $\Phi_{T1}$ and $\Phi_{T2}$ (790 K < ***T****< 1080K) is most appropriate. This is because the final temperature of the surface layer (***T****) is well below $T_c$ and the metastable metallic phase must be driven by the non-thermal action of the pump pulse.

To confirm the metallic character of the photo-induced phase in $NbO_2$ we present in figure 3(a-b) the transient THz conductivity $\Delta\sigma(\omega)$, for pump fluences of 15 mJ/cm$^2$ (***T**** ~ *960 K*) and 20.8 mJ/cm$^2$ (***T**** ~ *1210 K)* measured at different probe delays after the pump pulse. The finite conductivity response for 15 mJ/cm$^2$ (***T**** ~ *960 K*) is shown in figure 3(a), and attests its purely nonthermal character below $T_c$. To provide a physical description of the carrier properties, the transient complex conductivity $\Delta\sigma(\omega)$ of the $NbO_2$ film can be fitted using the Drude-Smith model given by, $\Delta\sigma(\omega) = \frac{\varepsilon_o \omega_p^2 \tau_o}{1-i\omega\tau_o}\left(1 + \frac{c}{1-i\omega\tau_o}\right)$, wherein $\varepsilon_o$, $\omega_p$, and $\tau_o$ represents the vacuum permittivity, the plasma frequency, and the carrier scattering time, respectively [21]. This model presents a good description of the ultrafast photo-induced IMT in several systems where there is a coexistence of metallic and insulating nanodomains [22, 23]. Here the fitting parameter ***c*** which is attributed to the persistence of the velocity after first scattering event can have values between -1 and 0, while the former value indicates full backscattering and/or carrier localization while the latter corresponds to the Drude model.

The electron localization in the metallic domains and the nonzero value of ***c*** points towards the backscattering at the domain walls or grain boundaries [17] and may also result from the Coulombic restoring force from the positively charged holes and the electrons. During the ultrafast phase transition, electrons are excited into the conduction band which results in the formation of the metallic domains in the insulating matrix. A mutual phase competition between the insulating and the metallic nano-domains will occur. The overall macroscopic character of the film is dictated by the growth and coalescence of the insulating or metallic domains. For $NbO_2$ the value of $c$ varies from -0.52 to -0.65 as the fluence is increased from 15 mJ/cm$^2$ to 20.8 mJ/cm$^2$ suggesting a more Drude like character for the former than for the latter. We find that $\Delta\sigma(\omega)$ decreases as a function of delay from (1 to 2.5 ps), which suggests that domains of the insulating phase grow and coalesce at the expense of metallic domains. This will result in a decrease of the carrier concentration as well as the overall conductivity, further suggesting the growth of the insulating domains [Table I]. The carrier concentration was calculated using $N = \varepsilon_0 m^* \omega_p^2 / e^2$, where the effective mass $m^*=12.7m_o$ [9], $m_o$ is the electron mass and the dc conductivity is computed using $\sigma_o = \varepsilon_0 \tau_o \omega_p^2 (1 + c)$. Our calculated values of $N$ are of the order of ($10^{19}$ cm$^{-3}$) lesser than the value of $N$ (~$10^{22}$ cm$^{-3}$) [9] when the sample is thermally heated above $T_c$. This is expected as we switch only 60nm depth of the film. However, there is only a marginal increase in the value of $N$ with pump fluence from 15 mJ/cm$^2$ to 20.8 mJ/cm$^2$, which suggests the effect of the matrix of insulating $NbO_2$ in capturing the photoexcited electrons via recombination limiting the overall conductivity and metallicity. The mobility was found using $\mu = q\tau_o(1 + c)/m^*$ and the values for a 1 ps probe delay were 2.72 cm$^2$/Vs for 15 mJ/cm$^2$ which increases to 2.9 cm$^2$/Vs for 20.8 mJ/cm$^2$. These values are moderately higher than the value of 0.5 cm$^2$/Vs reported for a crystalline $NbO_2$ sample at 600 K [9].

We now discuss the OPTP results on the $NbO_2$ film for longer probe delays and the observed non-thermal character of the photo-induced metallic state. In $NbO_2$ samples we do not find $\Delta T/T$ to increase with the delay time for several tens of ps, even for maximum fluence 24.7 mJ/cm$^2$ [figure 2(a)], which is in contrast to $VO_2$ where a slow increase in the magnitude of $\Delta T/T$ is often noted above threshold excitation at later delays [7,19]. This increase indicates that for our $NbO_2$ film the metallicity remains restricted to the region of the penetration depth of 60 nm in the $NbO_2$ film and beyond this thickness, deeper layers (with a thickness of 130 nm) remain robustly insulating. Consequently, this effect is also seen in the $\Delta\sigma(\omega)$, which exhibits only a marginal decrease between the measurement at a 1 ps probe delay near the maximum of the $\Delta T/T$ signal, and at a 2.5 ps delay as shown in figure 3(b) and Table I. This decrease suggests that the insulating lattice surrounding the metallic domains of the intermediate metallic state does not have the support of the structural transition or any kind of phonon triggered nucleation mechanism. Thus any photo-induced metallic domains cannot expand to deeper layers. Cocker et al. have noted the similar effect at low temperatures for $VO_2$ [7]. It was stressed that pump fluence could create the critical electron density necessary for the metallicity, however it could not trigger the critical 6 THz phonons essential to drive the complete structural transition [4,7]. In the case of $VO_2$, the ultrafast phase transition possesses an intermediate step with the formation of a metallic band. For example, a monoclinic metal- like phase was detected before the system undergoes a full structural transition [24-26]. To form the metallic state in $VO_2$, the photo-doping of the holes in the 3d valence band can also close the band gap, inducing the metallic phase without any structural phase transition [5]. In the case of $NbO_2$, we can envisage similar scenarios. Our measurement of a finite value for the transient THz conductivity well below $\Phi_{T2}$ in the *non-thermal regime* indicates that the critical density of electrons required to switch to the intermediate metallic state is similarly achieved for the $NbO_2$ film. Further, the larger threshold fluence necessary to initiate this transition to the metallic phase is presumably due to the presence of a robust background of insulating phase in $NbO_2$. Overall, this purely non-thermal IMT can be tentatively understood as the suppression of the on-site Coulomb repulsion and the collapse of the band gap as discussed previously within the scenarios developed for the ultrafast IMT in $VO_2$ [2,7,24-25].

In conclusion, we demonstrate the non-thermal nature of the ultrafast insulator-to-metal transition in epitaxial $NbO_2$ thin film and confirm the metallic character of the photo-induced phase by transient THz conductivity measurements. The high $T_c$ of $NbO_2$ results in a large range over which the ultrafast intermediate metallic phase can be decoupled from the thermally driven structural transition. These results indicate that $NbO_2$ represents an important test bed for an understanding of the photo-induced IMT in strongly correlated oxides. In particular, further time-resolved studies using structural probes such as x-ray or electron diffraction may provide important clues about the interplay of electronic and structural degrees of freedom in this interesting case of the purely non-thermal phase transition.

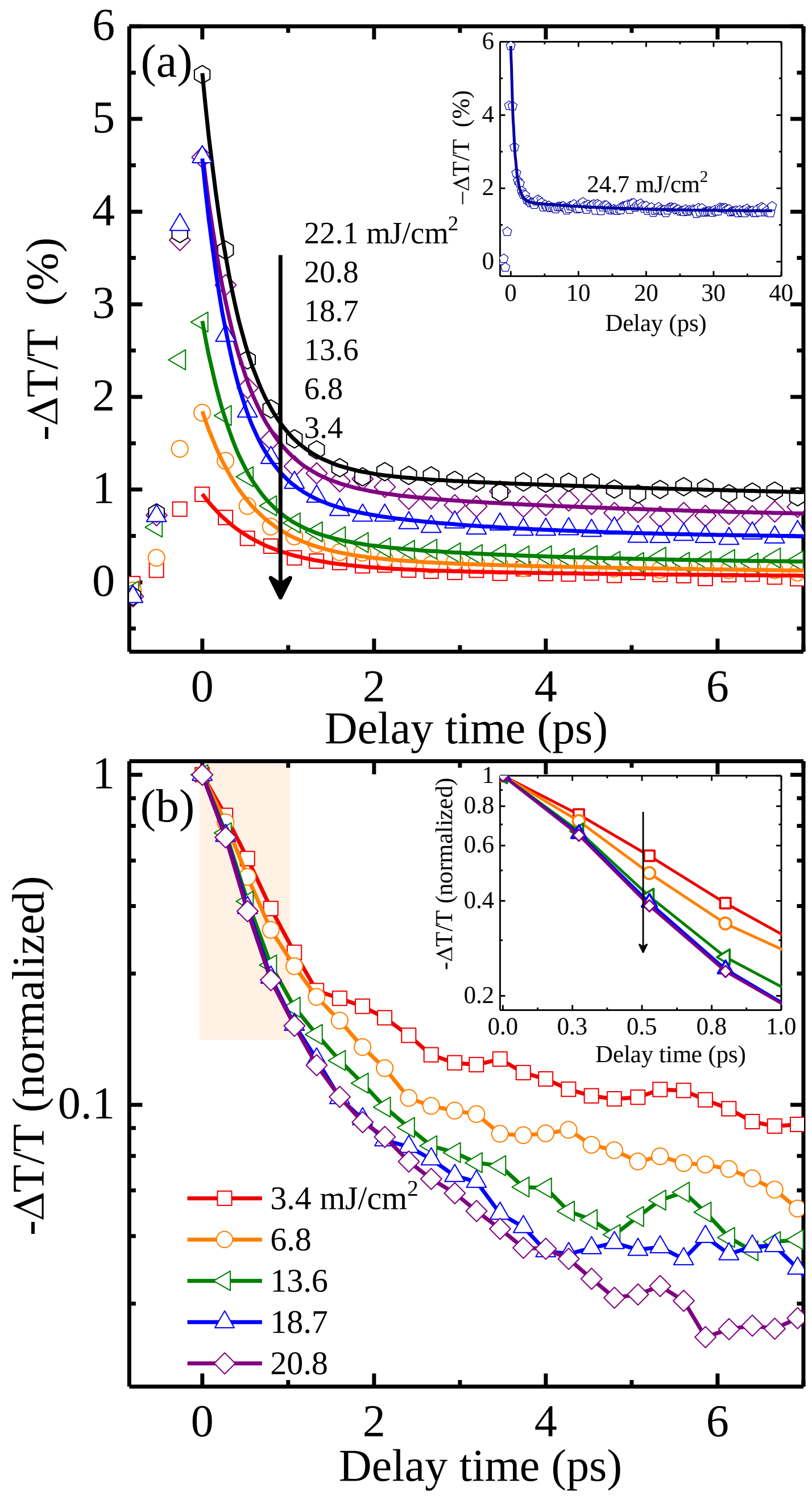

Fig. 1. (a) Spectrally integrated transient change of the THz transmission after excitation with a near-infrared (800 nm) pump pulse for various fluences at a substrate temperature of 295 K. The experimental data is fitted with bi-exponential functions and the inset depicts the signal at longer delay times (b) shows a semi-logarithmic plot of the normalized differential THz transmission versus delay for various fluences, inset depicts the enlarged view of the effect of fluence below 1 ps.

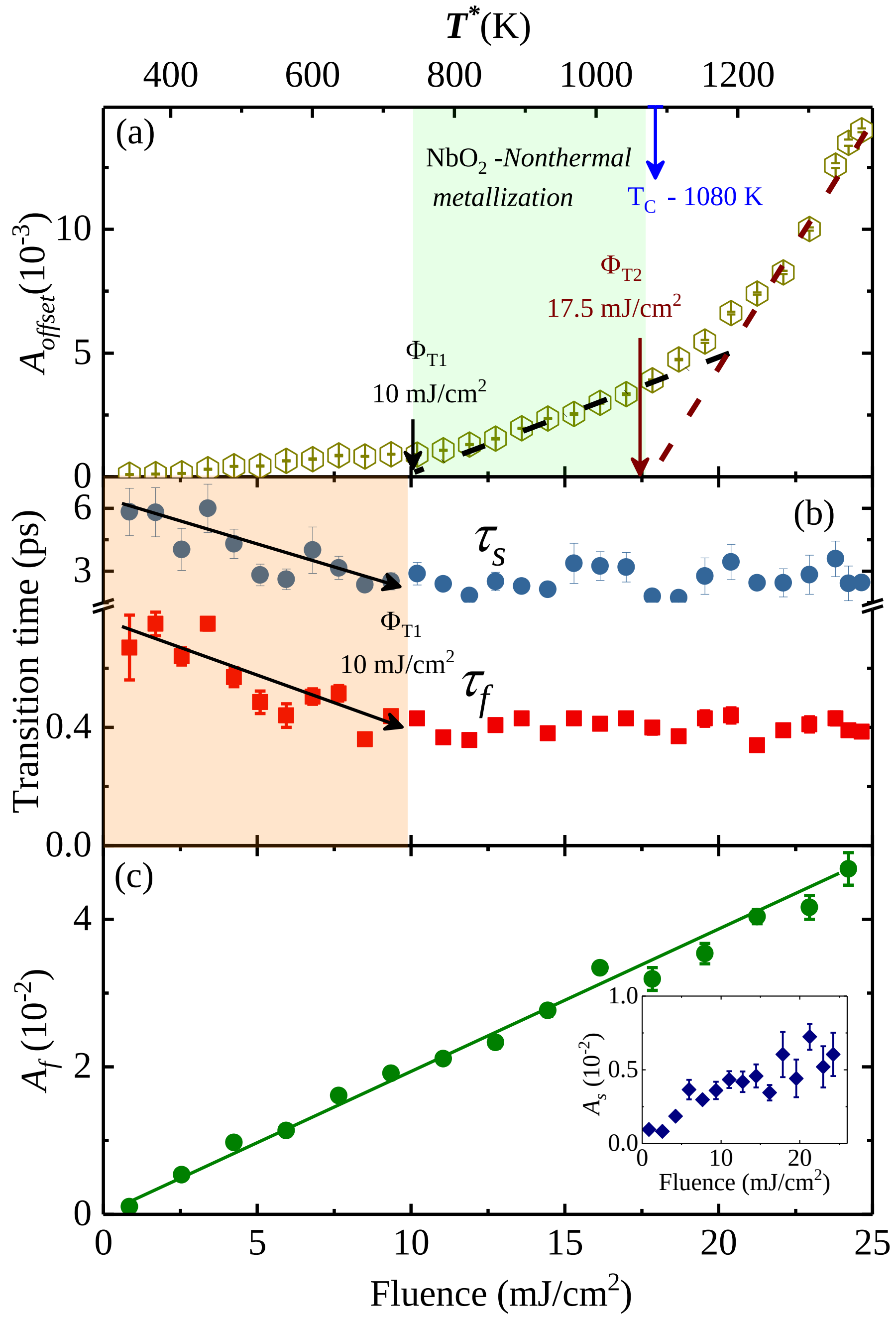

Fig. 2. (a) ΔT/T offset ($A_{offset}$) as a function of the pump fluence. The dashed lines show the linear extrapolations that give two characteristic threshold fluences $\Phi_{T1}$ and $\Phi_{T2}$, here ***T**** denotes the temperature of the photoexcited $NbO_2$ layer immediately after the pumping. (b) Variation of the fast dynamics captured by the time constant $\tau_f$, and the slow dynamics depicted by $\tau_s$ as a function of fluence. (c) The constant $A_f$ captures the amplitude of the fast dynamics as a function of fluence (straight green line is a guide to eye) and inset shows the variation of $A_s$ (depicting the amplitude of slow dynamics) as a function of fluence.

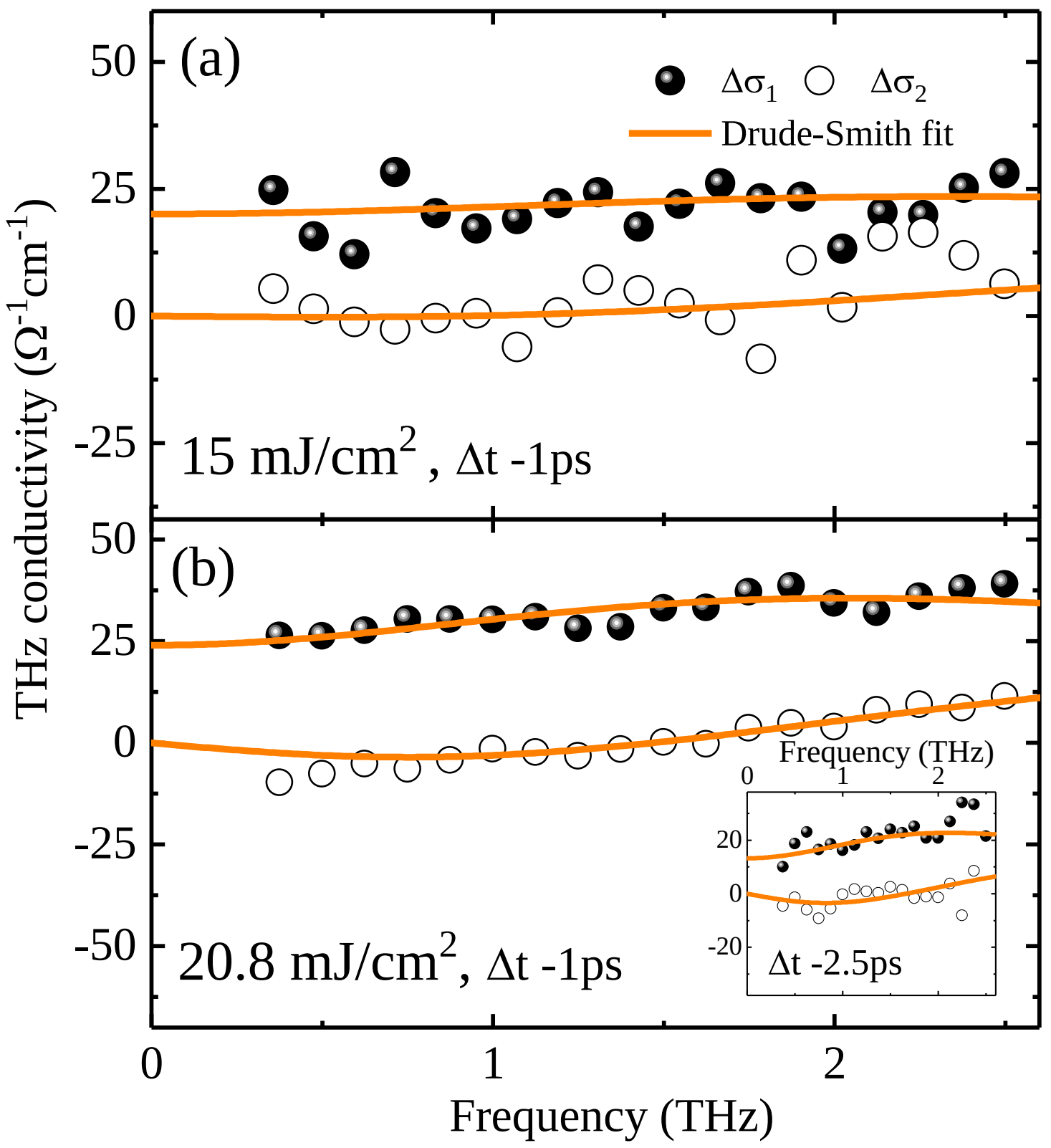

Fig. 3. (a) The transient THz complex conductivity spectra where $\Delta\sigma_1$ is the real and $\Delta\sigma_2$ is the imaginary part of the transient complex conductivity for different pump probe delays (Δt) after photoexcitation. Here solid lines represent fits using the Drude-Smith model.

**Table I**. The Drude-Smith fitting parameters obtained after fitting the transient optical conductivity data.

| **Fluence** | **Delay time (ps)** | $\omega_p$ **(THz)** | $\tau_o$ **(fs)** | ***c*** | ***N*** ***(*10$^{19}$ cm$^{-3}$)*** | $\sigma_o$ **($\Omega^{-1}$cm$^{-1}$)** |
|---|---|---|---|---|---|---|
| 15 mJ/cm$^2$ | 1 | 107 | 41 | -0.52±0.06 | 4.6 | 20 |
| 20.8 mJ/cm$^2$ | 1 | 115 | 60 | -0.65±0.02 | 5.3 | 25 |
| 20.8 mJ/cm$^2$ | 2.5 | 90 | 58 | -0.62±0.04 | 3.2 | 16 |

*Supplementary information*

# Non-thermal nature of photo-induced insulator-to-metal transition in $NbO_2$

Rakesh Rana[1], J. Michael Klopf[2], Jörg Grenzer[1], Harald Schneider[1], Manfred Helm[1,3], Alexej Pashkin[1]

[1]*Institute of Ion Beam Physics and Material Research, Helmholtz-Zentrum-Dresden-Rossendorf, 01328 Dresden, Germany*

[2]*Institute of Radiation Physics, Helmholtz-Zentrum Dresden-Rossendorf, 01328 Dresden, Germany*

[3]*Institute of Applied Physics, Technische Universität Dresden, 01062 Dresden, Germany*

## Characterization of the $NbO_2$ thin film

The epitaxial $NbO_2$ film was grown on (111) $MgAl_2O_4$ substrate using pulsed laser deposition technique. The deposition was carried out using a $NbO_2$ ceramic target, with a KrF laser at a fluence of 2J/cm$^2$ with a 2 Hz pulse repetition rate. The substrate was kept at a temperature of 650 °C; a deposition pressure of 2 mTorr with an $O_2$/Ar ratio of 7:93 [1] was used. The thickness of the grown $NbO_2$ film is 190 nm. The phase purity was confirmed using PanAlytical x-ray diffractometer (Empyrean). The x-ray measurements were carried out using Cu-$K_\alpha$ radiation; the diffractometers were equipped with a Göbel mirror on the source side. For the $2\theta$ - $\theta$ scan shown in Fig. S1(a) a small slit in front of a point detector was used.

Unpolarized Raman spectrum shown in Fig. S1(b) was collected using a Jobin-Yvon LabRAM HR 800 spectrometer under the excitation at a wavelength of 532 nm. The signal was collected through a 100× microscope objective. The Raman peak positions agree well with previously reported data [2] and are comparably sharp indicating absence of inhomogeneous broadening due to vacancies or impurities.

The reciprocal space maps were carried out using an area detector. Assuming a room temperature $I4_1/a$ unit cell a (110) oriented growth which would correspond to a (100) pseudorutile orientation can be proposed. The films have an out of plane lattice spacing of $d_{(220)}$ = 0.484 nm. The thin film exhibits a columnar structure, a rough estimate gives a coherent lateral size in the order of ~5nm and coherent vertical size of about ~15..20 nm, i.e., the film thickness is about 10 times larger than the coherent crystallite size. The columnar character of the grains in our $NbO_2$ film may explain the relatively large localization parameters of the in-plane electronic conductivity deduced from the Drude-Smith analysis.

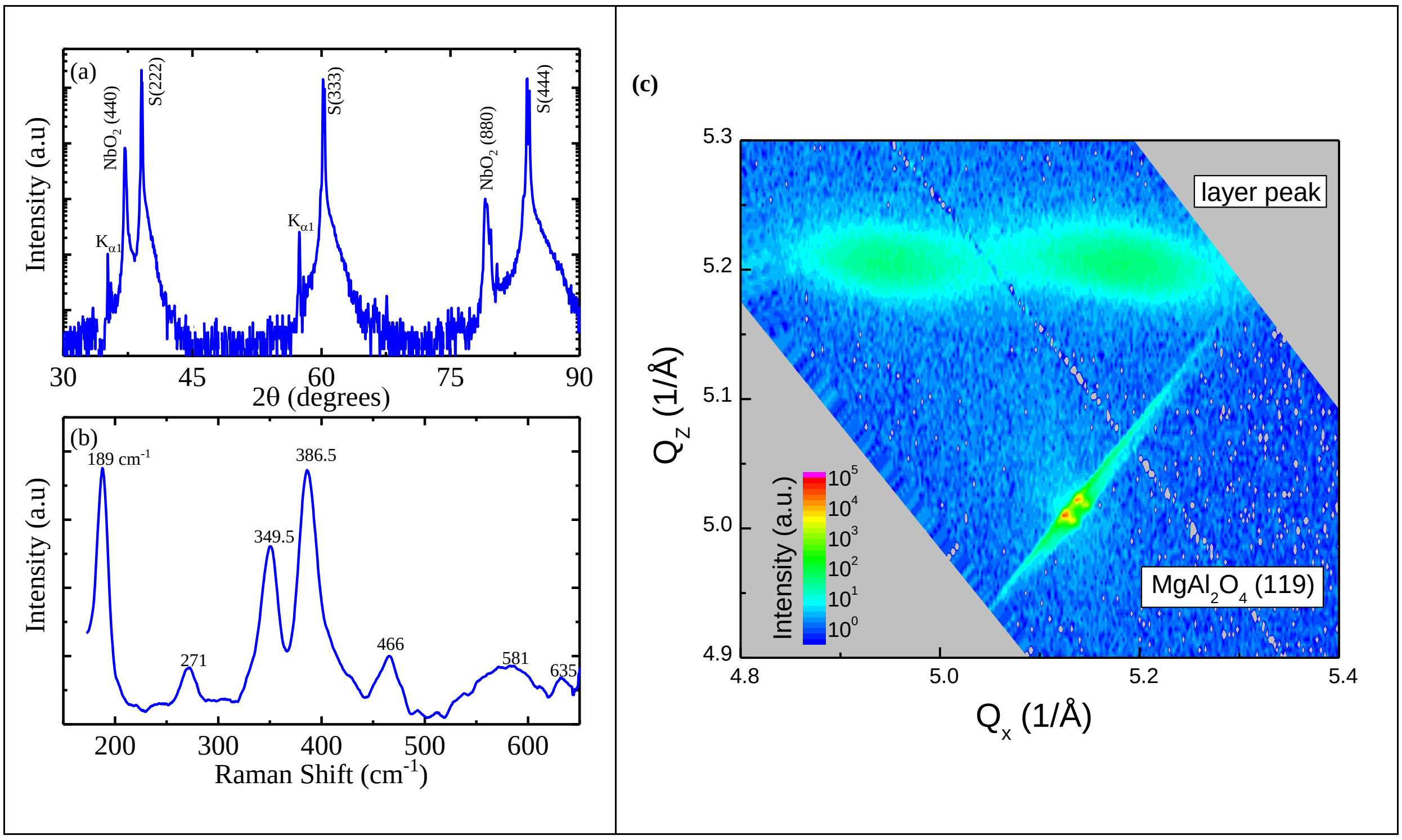

Fig. S1. (a) X-ray diffraction of the $NbO_2$ film on $MgAl_2O_4$ (111) substrate. The substrate peaks are denoted by S; (b) Raman spectrum of the $NbO_2$ thin film; (c) Reciprocal space map at the (119) substrate reflection. At larger $Q_x$ the layer peaks split off indicating that a domain structure is built up due to the lattice mismatch between the substrate and the thin film. All characterizations were performed at room temperature.